\documentclass[11pt,onecolumn,twoside]{IEEEtran}
\usepackage{cite}
\ifCLASSINFOpdf
 \usepackage[pdftex]{graphicx}
\else
 \usepackage[dvips]{graphicx}
\fi

\usepackage[cmex10]{amsmath}
\usepackage{amssymb}

\newtheorem{theorem}{Theorem}

\begin{document}
\title{Optimal Grouping for \\Group Minimax Hypothesis Testing}
\author{Kush~R.~Varshney and Lav~R.~Varshney
\thanks{Portions of the material in this paper were first presented in \cite{VarshneyV2011}.}
\thanks{K.~R.~Varshney is with the Business Analytics and Mathematical Sciences Department, IBM Thomas J.~Watson Research Center, Yorktown Heights, NY, 10598 USA (e-mail: krvarshn@us.ibm.com).}
\thanks{L.~R.~Varshney is with the Services Research Department, IBM Thomas J.~Watson Research Center, Yorktown Heights, NY, 10598 USA (e-mail: varshney@alum.mit.edu).}}

\maketitle

\begin{abstract}
Bayesian hypothesis testing and minimax hypothesis testing represent extreme instances of detection in which the prior probabilities of the hypotheses are either completely and precisely known, or are completely unknown.  Group minimax, also known as $\Gamma$-minimax, is a robust intermediary between Bayesian and minimax hypothesis testing that allows for coarse or partial advance knowledge of the hypothesis priors by using information on sets in which the prior lies.  Existing work on group minimax, however, does not consider the question of how to define the sets or groups of priors; it is assumed that the groups are given.  In this work, we propose a novel intermediate detection scheme formulated through the quantization of the space of prior probabilities that optimally determines groups and also representative priors within the groups.  We show that when viewed from a quantization perspective, group minimax amounts to determining centroids with a minimax Bayes risk error divergence distortion criterion: the appropriate Bregman divergence for this task.  Moreover, the optimal partitioning of the space of prior probabilities is a Bregman Voronoi diagram.  Together, the optimal grouping and representation points are an $\epsilon$-net with respect to Bayes risk error divergence, and permit a rate--distortion type asymptotic analysis of detection performance with the number of groups.  Examples of detecting signals corrupted by additive white Gaussian noise and of distinguishing exponentially-distributed signals are presented.  
\end{abstract}

\begin{IEEEkeywords}
Bayesian hypothesis testing, Bregman divergence, detection theory, minimax hypothesis testing, quantization, Stolarsky mean
\end{IEEEkeywords}

\IEEEpeerreviewmaketitle

\section{Introduction}
\label{sec:intro}
\IEEEPARstart{B}{ayesian} hypothesis testing and minimax hypothesis testing are signal detection formulations for when the prior probabilities of the hypotheses are perfectly and precisely known and for when the prior probabilities of the hypotheses are completely unknown, respectively \cite{VanTrees1968}.  Optimal performance in both settings is achieved by likelihood ratio tests with appropriately chosen thresholds.  Between these two edge cases, there is an entire set of likelihood ratio tests corresponding to a coarse knowledge of the prior probabilities; these intermediate formulations are explored in this work.

Formulations that lie between Bayesian and minimax hypothesis testing are known as group minimax or $\Gamma$-minimax and are of interest because it is difficult to obtain complete information about priors in many decision-making scenarios, but information about priors is also not completely lacking \cite{Robbins1950,Good1952,Savage1954,Robbins1964,BlumR1967,Vidakovic2000,Ruggeri2006}.  Group minimax detection formulations take partial information about priors as input and provide robustness against that partial information, in contrast to minimax hypothesis testing which provides robustness against complete lack of information.  Throughout the long history of group minimax statistical inference, the sets, groups, or $\Gamma$s in which the true priors lie are treated as inputs and are not optimized for detection or estimation performance.  In contrast, our work herein investigates the joint problem of optimizing the groupings within which to find a minimax-optimal representative prior as well as finding those priors for all groupings. 

We view the minimax test as one in which knowledge of prior probabilities has been quantized to a single cell encompassing the entire probability simplex, and the Bayesian test as one in which knowledge of prior probabilities has been quantized to an infinite number of cells that finely partition the probability simplex.  In the group minimax test, the prior probabilities are quantized to a finite number of cells.  The appropriate quantization distortion measure for prior probabilities of hypotheses is Bayes risk error \cite{VarshneyV2008b}, which is a Bregman divergence \cite{Varshney2011}.  However unlike standard quantization, we are interested in minimizing the maximum distortion rather than minimizing the average distortion \cite{GershoG1992,GrayN1998,GrafL2000}.  Thus we pursue minimax Bayes risk error quantization of prior probabilities \cite{VarshneyV2011}.

\IEEEpubidadjcol

Group minimax, which provides a means to consider intervals of prior belief rather than exact prior belief, is similar in spirit but differs in details to decision making based on interval-valued probability described in \cite{WolfensonF1982}.  There are also connections to representative prior distributions \cite{Hildreth1963}, the robust Bayesian viewpoint \cite{Berger1982,PericchiW1991}, and other areas of decision making in which robustness is desired \cite{BertsimasBC2011}.

To the best of our knowledge, there has been no previous work on the quantization of prior probabilities for hypothesis testing besides our own \cite{VarshneyV2008b,VarshneyRVG2011}.  Many interesting findings on average distortion clustering with Bregman divergences as the distortion criteria are reported in \cite{BanerjeeGW2005,BanerjeeMDG2005}, but we believe this is the first use of Bregman divergences in studying group minimax hypothesis testing.  Although studies and results in quantization theory typically focus on average distortion, maximum distortion does also appear occasionally, e.g.~\cite{ZhuH1999,GrafL2000,SarsharW2004,VenkitasubramaniamTS2006,Reznik2011}.  Such a minimax partitioning of a space is known as an $\epsilon$-net or $\epsilon$-covering \cite{KolmogorovT1961}.  

In investigating quantization for group minimax hypothesis testing, we derive centroid and nearest neighbor conditions for minimax Bayes risk error distortion and discuss how alternating application of these conditions leads to a locally optimal quantizer.  We provide direct derivations in the binary detection case and specialize elegant results from the Bregman divergence literature in the general case.  Minimax centroid conditions for Bregman divergences are derived in \cite{NockN2005}.  The problem of finding the optimal nearest neighbor cell boundaries for a given set of samples, also known as a Voronoi diagram, is addressed for Bregman divergences in \cite{NielsenBN2007,BoissonnatNN2010}.  Advantages of the direct derivations for the binary setting include direct geometric insights, as well as closed-form expressions.

As a further contribution similar in style to rate--distortion theory \cite{DonohoVDD1998}, we present asymptotic results on detection error as the partiality of information about the prior goes from the minimax hypothesis testing case to the Bayesian hypothesis testing case.  We also present a few examples of group minimax detection with different likelihood models.

The rest of the paper is organized in the following manner.  First in Section~\ref{sec:prelim}, we set forth notation and briefly provide background on Bayesian, minimax and group minimax detection, along with Bayes risk error divergence.  We formulate a quantization problem to find optimal groupings for group minimax detection in Section~\ref{sec:quant}.  Section~\ref{sec:optimality} derives the nearest neighbor and centroid optimality conditions for the proposed quantization problem in both the binary and $M$-ary cases.  We analyze the rate--distortion behavior of the groups in Section~\ref{sec:analysis}.  Two examples are presented in Section~\ref{sec:examples} to provide intuition.  Section~\ref{sec:conclusion} provides a summary of the contributions and concludes.  

\section{Preliminaries}
\label{sec:prelim}

The detection or hypothesis testing problem is the task of accurately determining which of $M$ classes a noisy signal instance belongs to.  In the binary ($M = 2$) case, this task is often determining the presence or absence of a target based on a measurement observed through noise.  In this section we first discuss binary hypothesis testing and then we consider $M$-ary hypothesis testing for $M > 2$.  Finally we present the definition of the Bayes risk error divergence, a quantification of detection performance degradation.

\subsection{Binary Decisions}
\label{sec:prelim:binary}

Consider the binary hypothesis testing problem.   There are two hypotheses $h_0$ and $h_1$ with prior probabilities $p_0 = \Pr[H=h_0]$ and $p_1 = \Pr[H=h_1] = 1 - p_0$, and a noisy observation $Y$ governed by likelihood functions $f_{Y|H}(y|H = h_0)$ and $f_{Y|H}(y|H = h_1)$.  A decision rule $\hat{h}(y)$ that uniquely maps every possible $y$ to either $h_0$ or $h_1$ is to be determined.  There are two types of error probabilities: 
\begin{align*}
p_E^{\text{I}} &= \Pr[\hat{h}(Y)=h_1 | H=h_0] \mbox{, and} \\ 
p_E^{\text{II}} &= \Pr[\hat{h}(Y)=h_0 | H=h_1]\mbox{.}
\end{align*}

Minimizing weighted error, the optimization criterion for the decision rule $\hat{y}(y)$ is the Bayes risk:
\begin{equation}
J = c_{10}p_0p_E^{\text{I}} + c_{01}(1-p_0)p_E^{\text{II}}\mbox{,}
\label{eq:bayesrisk}
\end{equation} 	
where $c_{10}$ is the cost of the first type of error and $c_{01}$ is the cost of the second type of error.  The decision rule that optimizes \eqref{eq:bayesrisk} is the following likelihood ratio test \cite{VanTrees1968}:
\begin{equation}
\frac{f_{Y|H}(y|H = h_1)}{f_{Y|H}(y|H = h_0)} \mathop{\gtreqless}^{\hat{h}(y)=h_1}_{\hat{h}(y)=h_0} \frac{p_0c_{10}}{(1-p_0)c_{01}}\mbox{.}
\label{eq:lrt}
\end{equation}
The prior probability $p_0$ appears on the right side of the rule in the threshold.  Since the prior probability is part of the specification of the Bayes-optimal decision rule, the error probabilities $p_E^{\text{I}}$ and $p_E^{\text{II}}$ are functions of the prior probability.  Thus we may write the Bayes risk as a function of $p_0$:
\begin{equation}
J(p_0) = c_{10}p_0p_E^{\text{I}}(p_0) + c_{01}(1-p_0)p_E^{\text{II}}(p_0)\mbox{,}
\label{eq:bayesrisk_f}
\end{equation} 	
The function $J(p_0)$ is zero at the points $p_0=0$ and $p_0=1$ and is positive-valued, strictly concave, and continuous in the interval $(0,1)$ \cite{Wijsman1970}.  Under deterministic decision rules, $J(p_0)$ is differentiable everywhere.

The Bayesian hypothesis testing threshold on the right side of \eqref{eq:lrt} relies on the true prior probability $p_0$, but as discussed in Section~\ref{sec:intro}, this value may not be known precisely.  When the true prior probability is $p_0$, but the threshold in $\hat{h}(y)$ uses some other \emph{decision weight} $a$, there is mismatch.  The Bayes risk of the decision rule with threshold 
\[
\frac{ac_{10}}{(1-a)c_{01}}
\] 
is:
\begin{equation}
\label{eq:bayesrisk_mismatch}
J(p_0,a) = c_{10}p_0p_E^{\text{I}}(a) + c_{01}(1-p_0)p_E^{\text{II}}(a)\mbox{.}
\end{equation}

\begin{figure}
  \centering
  \includegraphics[width=0.49\textwidth]{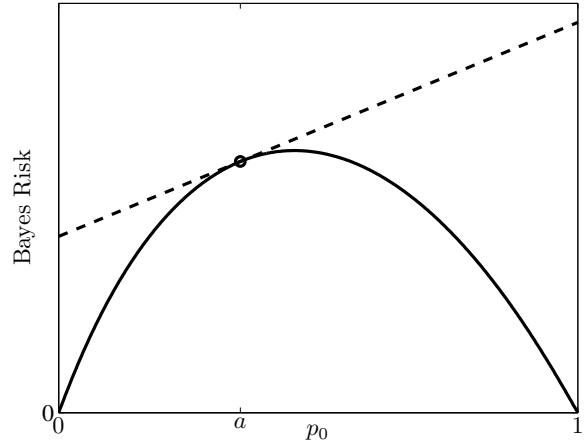}
  \caption{Example $J(p_0)$ (solid curve) and $J(p_0,a)$ (dashed line).}
  \label{fig:bayesriskex}
\end{figure}

The function $J(p_0,a)$ is a linear function of $p_0$ with slope $(c_{10}p_E^{\text{I}}(a) - c_{01}p_E^{\text{II}}(a))$ and intercept $c_{01}p_E^{\text{II}}(a)$.  The function $J(p_0,a)$ is tangent to $J(p_0)$ at $a$ and $J(p_0,p_0) = J(p_0)$.  By the point-slope formula of lines, the mismatched Bayes risk is also:
\begin{equation}
\label{eq:bayesrisk_mismatch_ps}
J(p_0,a) = J(a) + (p_0 - a)J'(a)
\end{equation}
when $J$ is differentiable.
An example of how $J(p_0)$ and $J(p_0,a)$ are related is shown in Fig.~\ref{fig:bayesriskex}.

The minimax hypothesis testing threshold is determined by finding the decision weight $a$ that minimizes the worst-case $J(p_0,a)$, that is:
\begin{equation}
a_{\rm{minimax}}^* = \arg\min_{a}\max_{p_0} J(p_0,a)\mbox{.}
\label{eq:minimaxpoint}
\end{equation}
Under equivalent notation, the optimal Bayesian decision weight is $a_{\rm{Bayesian}}^* = p_0$.  In Bayesian hypothesis testing, the decision weight $a$ continually changes with $p_0$, whereas in minimax hypothesis testing, there is a single decision weight $a$ for all $p_0$.  

\subsection{$M$-ary Decisions}
\label{sec:prelim:mary}

The basics from the binary case carry over to the $M$-ary case.  With $M$ hypotheses, there are $M$ prior probabilities $p_i > 0, i = 0, \ldots,M - 1$ such that $\sum_i p_i = 1$.  The collection of priors is denoted by the vector $\bf{p}$, which is an element of the $M$-ary probability simplex.  There is also an $M \times M$ matrix of costs $c_{ij}$.  The detection rule $\hat{h}(y)$ in the $M$-ary case uses ratios of priors and costs in an analogous manner to the likelihood ratio test \eqref{eq:lrt}.  The Bayes risk function is now
\begin{equation}
J({\bf p}) = \sum_{i=0}^{M-1}\sum_{j=0}^{M-1} c_{ij} p_j  \Pr[\hat{h}(Y,{\bf p}) = h_i | H = h_j] \mbox{.}
\label{eq:bayesrisk_vec}
\end{equation}

With a vector-valued decision weight $\bf{a}$, the mismatched Bayes risk function is 
\begin{equation}
J({\bf p},{\bf a}) = J({\bf a}) + ({\bf p} - {\bf a})^T \nabla J({\bf a}) 
\label{eq:bayesrisk_mismatch_vec}
\end{equation}
when $J$ is differentiable.
In the $M$-ary case, as in the binary case, $\mathbf{a}_{\rm{Bayesian}}^* = \mathbf{p}$ and
\begin{equation}	
	\label{eq:minimaxpointM}
	\mathbf{a}_{\rm{minimax}}^* = \arg\min_{\mathbf{a}}\max_{\mathbf{p}} J(\mathbf{p},\mathbf{a}).
\end{equation}

\subsection{Bayes Risk Error Divergence}
\label{sec:prelim:bre}

The Bayes risk $J(\mathbf{p})$ represents the performance of the best possible decision making under uncertainty, whereas the mismatched Bayes risk $J(\mathbf{p},\mathbf{a})$ represents the degraded decision-making performance due to the decision weight $\mathbf{a}$.  Thus, we may quantify the degradation or distortion in detection performance using the difference:
\begin{align}
	\label{eq:bre}
	d(\mathbf{p}\|\mathbf{a}) &= J(\mathbf{p},\mathbf{a}) - J(\mathbf{p}) \\
	\label{eq:bregrad}
	&= -J(\mathbf{p}) + J(\mathbf{a}) + (\mathbf{p} - \mathbf{a})^T \nabla J(\mathbf{a}).
\end{align}
This difference is a Bregman divergence termed \emph{Bayes risk error divergence} generated by the convex function $-J(\mathbf{p})$ over a convex domain (the $M$-ary probability simplex) \cite{VarshneyV2008b,Varshney2011}.

\section{Minimax Bayes Risk Error Quantization}
\label{sec:quant}

Having described a divergence that quantifies loss in detection performance due to a mismatched decision weight, in this section we describe how that divergence can be utilized within a scalar or vector quantization framework to yield not only the optimal minimax representation point for a given set of priors (the typical group minimax problem), but also the optimal groupings for the group minimax scenario.

\subsection{Quantization for Group Minimax Grouping}
\label{sec:quant:group}

The space of all possible decision weights and the space of all true prior probability vectors is the $M$-ary probability simplex.  As discussed in Section~\ref{sec:prelim}, in the Bayesian case, the decision weight $\mathbf{a}$ changes continuously with the true prior probability vector $\mathbf{p}$ of the detection problem, so that the Bayes risk error divergence $d(\mathbf{p}\|\mathbf{a}) = 0$ for all detection problems.  Denoting the mapping from true prior probability $\mathbf{p}$ to decision weight $\mathbf{a}$ as $q(\mathbf{p}) = \mathbf{a}$, this function is the identity function $q(\mathbf{p}) = \mathbf{p}$ in the Bayesian case and has the entire $M$-ary probability simplex as its range.

On the other hand in the minimax case, there is a single decision weight $\mathbf{a}_{\rm{minimax}}^*$ for all detection problems and $d(\mathbf{p}\|\mathbf{a}_{\rm{minimax}}^*) > 0$ for all problems except the one problem in which, by chance, the minimax decision weight is the true prior probability.  Here, the mapping from true prior probability to decision weight is a function whose range contains a single point: $q(\mathbf{p}) = \mathbf{a}_{\rm{minimax}}^*$.  

As discussed in Section~\ref{sec:intro}, it may be that the true prior probability and thus the Bayesian decision weight is not exactly known.  It may also be, however, that there is some partial information, and thus we need not restrict ourselves to just one decision weight for all detection problems but may have $K$ different decision weights.  Therefore, we would like to consider functions $q(\cdot)$ whose range is a finite set of $K$ decision weights $\{\mathbf{a}_1,\ldots,\mathbf{a}_K\}$.  With such a range, there are $K$ true prior probabilities for which there is no degradation in detection performance, i.e.~$d(\mathbf{p}\|q(\mathbf{p})) = 0$.  The function $q(\cdot)$ depends discontinuously on $\mathbf{p}$ such that $q(\mathbf{p}) = \mathbf{a}_k$ for all $\mathbf{p} \in \mathcal{Q}_k$, $k=1,\ldots,K$.  Such a function is a quantizer.  The remaining question for the proposed optimal grouping for group minimax hypothesis testing is determining the decision weights $\mathbf{a}_k$ and the quantization cells $\mathcal{Q}_k$.

\subsection{Minimax Bayes Risk Error Quantization Criterion}
\label{sec:quant:xbre}

Robustness is the motivation for both minimax hypothesis testing and group minimax hypothesis testing.  We take maximum Bayes risk error divergence as the objective for finding the decision weights and the quantization cells, resulting in the following minimax quantizer design problem:
\begin{equation}
\label{eq:multilevel_vec}
	q_K^* = \arg\min_{q_K}\max_{ {\bf p}} d({\bf p}\|q_K({\bf p})),
\end{equation}
where $q_K(\cdot)$ is a quantizer function with $K \ge 1$ cells and decision weights, and $K$ is a fixed parameter.  Operationally, knowing in advance that the true prior probability $\mathbf{p}$ falls in cell $\mathcal{Q}_k$ indicates that the decision weight $\mathbf{a}_k$ be used in setting the threshold.

\begin{figure}
  \centering
  \includegraphics[width=0.49\textwidth]{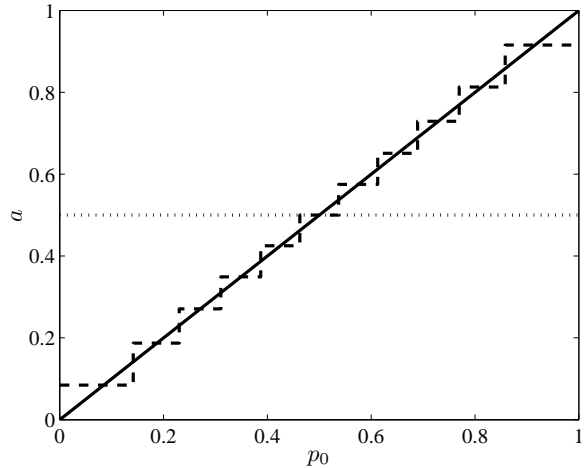}
  \caption{Example decision weight $a$ as a function of prior probability $p_0$ for Bayesian (solid), group minimax with eleven groups (dashed), and minimax (dotted) binary hypothesis testing.}
  \label{fig:dweight}
\end{figure}

In the $K = 1$ case, it is straightforward to show that the decision weight of $q_1^*(\cdot)$ equals the minimax hypothesis testing value $\mathbf{a}_{\text{minimax}}^*$, and occurs at the peak of $J(\mathbf{p})$.  However for $K>1$, the decision weight $\mathbf{a}_k$ within a cell $\mathcal{Q}_k$ is not the point that minimizes the maximum mismatched Bayes risk $J(\mathbf{p},\mathbf{a})$; rather it is the point that minimizes the maximum Bayes risk error divergence $d(\mathbf{p}\|\mathbf{a})$.  An example of the decision weight as a function of prior probability is shown in Fig.~\ref{fig:dweight}.  

In this section, we have defined an approach for optimal grouping for group minimax hypothesis testing.  This formulation reduces to the two extreme hypothesis testing methodologies: minimax at $K=1$ and Bayesian as $K\to\infty$.  

\section{Optimality Conditions}
\label{sec:optimality}

This section develops necessary conditions for optimality of a quantizer for the probability simplex under the minimax criterion \eqref{eq:multilevel_vec} defined in Section~\ref{sec:quant}, first in the scalar quantization (binary hypothesis testing) setting and then in the vector quantization ($M$-ary hypothesis testing) setting.  We find a centroid condition to locally optimize decision weights when the quantization cells are fixed.  Then we find a nearest neighbor condition to locally optimize the quantization cells with decision weights fixed.  Optimal quantizers can be found by alternately applying the nearest neighbor and centroid conditions through a version of the iterative Lloyd--Max algorithm \cite{GershoG1992,GrafL2000,NielsenBN2007}.  We provide direct derivations for the binary case and specialize more general Bregman divergence results for the $M$-ary case.

\subsection{Binary Hypothesis Testing Centroid Condition}
\label{sec:optimality:2centroid}
	
The $K$-cell scalar quantizer function $q_K(p_0)$ in the binary hypothesis testing problem has cell notation as follows.  The probability simplex $[0,1]$ is partitioned into $K$ intervals $\mathcal{Q}_1 = [0,b_1]$, $\mathcal{Q}_2 = (b_1,b_2]$, $\mathcal{Q}_3 = (b_2,b_3]$, \ldots, $\mathcal{Q}_K = (b_{K-1},1]$.  Within a fixed scalar quantization cell $\mathcal{Q}_k$ with boundaries $b_{k-1}$ and $b_{k}$,\footnote{Since $d(p_0\|a)$ increases monotonically with the absolute error, we can observe the convexity of the nearest neighbor cell; consequently each cell must consist of a single interval, cf.~\cite[Lemma 6.2.1]{GershoG1992}.} we want an expression for the optimal decision weight: \begin{equation}
\label{eq:centroidopt}
	a_k = \arg\min_{a\in\mathcal{Q}_k}\max_{p_0\in\mathcal{Q}_k} d(p_0\|a).
\end{equation}

\begin{theorem}
In the binary hypothesis testing problem with deterministic likelihood ratio test decision rules, the minimax Bayes risk error divergence optimal decision weight $a_k$ satisfies:
\begin{equation}
\label{eq:ak}
	J'(a_k) = \frac{J(b_k) - J(b_{k-1})}{b_k - b_{k-1}}.
\end{equation}
\end{theorem}
\begin{IEEEproof}
Let us first focus on the inner maximization in \eqref{eq:centroidopt}.  In the binary hypothesis testing case, 
\begin{equation}
\label{eq:line}
	d(p_0\|a) = -J(p_0) + J(a) + (p_0 - a)J'(a),
\end{equation}
from which we see that the second derivative of $d(p_0\|a)$ with respect to $p_0$ is $-J''(p_0)$, which is greater than zero due to the strict concavity of $J(p_0)$.  Thus, $d(p_0\|a)$ has no local maxima in the interior of $\mathcal{Q}_k$; the maximum occurs at an endpoint: $b_k$ or $b_{k-1}$.  Consequently,
\begin{align}
\label{eq:maxoftwo}
	\max_{p_0\in\mathcal{Q}_k} d(p_0\|a) &= \max\{d(b_k\|a),d(b_{k-1}\|a)\}\\
		&= \frac{d(b_{k-1}\|a) + d(b_k\|a) + |d(b_{k-1}\|a) - d(b_k\|a)|}{2}.
\end{align}
Substituting \eqref{eq:line} into \eqref{eq:maxoftwo} and simplifying, we find that \eqref{eq:maxoftwo} equals
\begin{equation}
\label{eq:minobjective}
	\frac{(b_{k-1} + b_k - 2a)J'(a) - J(b_{k-1}) - J(b_k) + 2J(a)}{2} + \frac{|(b_{k-1} - b_k)J'(a) - J(b_{k-1}) + J(b_k)|}{2},
\end{equation}
which is to be minimized with respect to $a\in\mathcal{Q}_k$.  

Due to the absolute value function, there are two cases to consider: 
\begin{enumerate}
	\item $(b_{k-1} - b_k)J'(a) - J(b_{k-1}) + J(b_k) \le 0$ and 
	\item $(b_{k-1} - b_k)J'(a) - J(b_{k-1}) + J(b_k) > 0$.
\end{enumerate}
Due to the concavity of the Bayes risk function, $J'(a)$ is monotonically decreasing.  Therefore, since $(b_{k-1} - b_k)$ is negative, $(b_{k-1} - b_k)J'(a) - J(b_{k-1}) + J(b_k)$ is a monotonically increasing function of $a$.  Consequently the two cases of the absolute value correspond to the intervals $(b_{k-1},a^\dagger]$ for case 1 and $(a^\dagger,b_k]$ for case 2, where $a^\dagger$ satisfies:
\begin{equation}
\label{eq:adagger}
	(b_{k-1} - b_k)J'(a^\dagger) - J(b_{k-1}) + J(b_k) = 0.
\end{equation}

In the first case, \eqref{eq:minobjective} simplifies to:
\begin{displaymath}
	(b_k - a)J'(a) + J(a) - J(b_k) 
\end{displaymath}
with derivative with respect to $a$:
\begin{displaymath}
	(b_k - a)J''(a),
\end{displaymath}
which is less than zero because $(b_k - a) > 0$ and $J''(a) < 0$ due to Bayes risk concavity.  Thus the minimization objective is monotonically decreasing in the first case.

In the second case, \eqref{eq:minobjective} simplifies to:
\begin{displaymath}
	(b_{k-1} - a)J'(a) + J(a) - J(b_{k-1}),
\end{displaymath}
which has derivative with respect to $a$:
\begin{displaymath}
	(b_{k-1} - a)J''(a),
\end{displaymath}
which is greater than zero because $(b_{k-1} - a) < 0$ and $J''(a) < 0$.  In the second case, the minimization objective is monotonically increasing.

Since \eqref{eq:minobjective} is decreasing over $(b_{k-1},a^\dagger]$ and increasing over $(a^\dagger,b_k]$, it is minimized at $a^\dagger$.  Therefore $a_k = a^\dagger$.  The decision weight satisfies \eqref{eq:adagger}.  This is equivalently the slope matching condition \eqref{eq:ak} given in the statement of the theorem.
\end{IEEEproof}

This minimax centroid is a Stolarsky mean \cite{Stolarsky1975}; the Stolarsky mean of $u$ and $v$ is in general:
\begin{displaymath}
	F'^{-1}\left(\frac{F(u)-F(v)}{u-v}\right)
\end{displaymath}
for any reasonable function $F(\cdot)$.

\subsection{Binary Hypothesis Testing Nearest Neighbor Condition}
\label{sec:optimality:2neighbor}

In the binary hypothesis testing nearest neighbor condition, we are to find the cell boundary $b_k$ given the decision weights $a_k$ and $a_{k+1}$.  
\begin{theorem}
In the binary hypothesis testing problem with deterministic likelihood ratio test decision rules, the minimax Bayes risk error divergence optimal cell boundary $b_k$ is:
\begin{equation}
\label{eq:bk}
	b_k = \frac{a_{k+1}J'(a_{k+1}) - a_kJ'(a_k) - (J(a_{k+1}) - J(a_k))}{J'(a_{k+1}) - J'(a_k)}.
\end{equation}
\end{theorem}
\begin{IEEEproof}
As discussed in Section~\ref{sec:optimality:2centroid}, the maximum Bayes risk error divergence within a cell occurs at the cell boundary.  Therefore, we would like to minimize the Bayes risk error divergence at the cell boundary.  

Specifically, $b_k$ should be chosen to minimize the maximum of $d(b_k\|a_k)$ and $d(b_k\|a_{k+1})$.  At a given potential boundary point $b$, the $J(b)$ term is the same in both $d(b\|a_k)$ and $d(b\|a_{k+1})$, so only $J(b,a_k)$ and $J(b,a_{k+1})$ need be considered.  Due to the geometry of the problem, $b_k$ should be the abscissa of the point at which the lines $J(p_0,a_k)$ and $J(p_0,a_{k+1})$ intersect.  Working with the definitions of $J(p_0,a_k)$ and $J(p_0,a_{k+1})$, we find the point of intersection to be \eqref{eq:bk}. 
\end{IEEEproof}

The cell boundary is the tangent line mean of the decision weights \cite{DietelG2003}.  The nearest neighbor condition for minimax Bayes risk error quantization is the same as that for minimum mean Bayes risk error quantization \cite{VarshneyV2008b}.

\subsection{$M$-ary Hypothesis Testing Nearest Neighbor Condition}
\label{sec:optimality:Mneighbor}

We found the nearest neighbor condition over the binary simplex, i.e., the line segment between zero and one, in Section~\ref{sec:optimality:2neighbor}.  In that case, the cell boundaries were simply two points on the line.  The situation is slightly more complicated notationally in the $M$-ary detection task because of the increased dimensionality.  Let us define the $M$-ary probability simplex as follows:
\begin{equation}
\label{eq:Marysimplex}
	\mathcal{P}_M = \left\{ \boldsymbol{\pi} \in \mathbb{R}_{+}^{M-1} \biggm| \sum_{i=1}^{M-1} \pi_i \le 1 \right\}.
\end{equation}
Now in specifying the nearest neighbor condition, we assume that the decision weights $\{\mathbf{a}_1,\ldots,\mathbf{a}_K\}$ are fixed.  We denote the set of points in $\mathcal{P}_M$ that are equidistant according to Bayes risk error divergence from $\mathbf{a}_k$ and $\mathbf{a}_{k+1}$ as $\mathcal{B}_{k,k+1}$, such that
\begin{equation}
\label{eq:Bset}
	\mathcal{B}_{k,k+1} = \left\{ \boldsymbol{\pi} \in \mathcal{P}_M \mid d(\boldsymbol{\pi}\|\mathbf{a}_k) = d(\boldsymbol{\pi}\|\mathbf{a}_{k+1}) \right\}.
\end{equation}

We show that this bisector $\mathcal{B}_{k,k+1}$ between the two decision weights $\mathbf{a}_k$ and $\mathbf{a}_{k+1}$ is a hyperplane in $\mathcal{P}_M$.
\begin{theorem}
In the $M$-ary hypothesis testing problem with deterministic likelihood ratio test decision rules, the Bayes risk error divergence bisector $\mathcal{B}_{k,k+1}$ satisfies the hyperplane equation:
\begin{equation}
\label{eq:B_kkp1}
	\mathcal{B}_{k,k+1} = \left\{\boldsymbol{\pi} \in \mathcal{P}_M \mid \boldsymbol{\pi}^T\left(\nabla J(\mathbf{a}_{k+1}) - \nabla J(\mathbf{a}_{k})\right) = \mathbf{a}_{k+1}^T\nabla J(\mathbf{a}_{k+1}) - \mathbf{a}_k^T\nabla J(\mathbf{a}_k) - \left(J(\mathbf{a}_{k+1}) - J(\mathbf{a}_k)\right)\right\}.
\end{equation}
\end{theorem}
\begin{IEEEproof}
The result follows by specializing \cite[Lemma 4]{NielsenBN2007}, which applies to all Bregman divergences, to Bayes risk error divergence.
\end{IEEEproof}

It is easy to see that we recover the binary boundary expression for $b_k$ \eqref{eq:bk} when we set $M = 2$ in \eqref{eq:B_kkp1}.

In the binary case, the boundary point bisectors fully specify the quantization cells $\mathcal{Q}_k$, but in the $M$-ary case, we must go one step further.  In particular, the quantization cells are defined as follows:
\begin{equation}
\label{eq:MQk}
	\mathcal{Q}_k = \left\{\boldsymbol{\pi} \in \mathcal{P}_M \mid d(\boldsymbol{\pi}\|\mathbf{a}_k) \le d(\boldsymbol{\pi}\|\mathbf{a}_{k'}), \, k' \neq k\right\}.
\end{equation}
Moreover, as discussed in \cite{NielsenBN2007}, the cell is a convex polyhedron, which is delineated by the intersection of the bisectors between its decision weight and all other cell decision weights.  The set of all minimax quantization cells is the Voronoi diagram of the simplex with the set of fixed decision weights as seeds.  If we write the half space induced by $\mathcal{B}_{k,k'}$ such that it contains $\mathbf{a}_k$ and is restricted to $\mathcal{P}_M$ as $\mathcal{H}_{k,k'}$, then
\begin{equation}
\label{eq:MQkint}
	\mathcal{Q}_k = \bigcap_{k' \neq k} \mathcal{H}_{k,k'}.
\end{equation}

Let $v_k$ be the number of vertices of cell $\mathcal{Q}_k$.  Each $\mathcal{Q}_k$ has at most $(K-1) + (M-1)$ faces and at most $\binom{K-1}{M-2} + M$ vertices, i.e.~$v_k \le \binom{K-1}{M-2} + M$.  (The constant additive terms that correspond to the dimension of the space are due to intersections with the simplex boundary.)  Moreover, in the same way that the maximum Bayes risk error divergence within a cell occurs at the cell boundary in the binary case, the maximum Bayes risk error divergence occurs at one of the finite $v_k$ vertices in the $M$-ary case \cite[Lemma 12]{NielsenBN2007}.

\subsection{$M$-ary Hypothesis Testing Centroid Condition}
\label{sec:optimality:Mcentroid}

In Section~\ref{sec:optimality:2centroid}, we found the minimax Bayes risk error centroid condition in the binary detection case through an explicit calculation that made use of convexity properties of the Bayes risk function.  Here we find the centroid condition in general for $M$-ary detection, by adapting the minimax centroid results of general Bregman divergences found in \cite{NockN2005}.  

In deriving the centroid condition, the cell $\mathcal{Q}_k$ and its $v_k$ vertices are fixed.  Since we know the maximum divergence occurs at a vertex, we only examine the vertices of $\mathcal{Q}_k$ in order to find the minimax-optimal decision weight within the cell.  Let the vertices of $\mathcal{Q}_k$ be denoted $\left\{\mathbf{b}_{k,1},\ldots,\mathbf{b}_{k,v_k}\right\}$.  The optimal decision weight is a functional mean of the vertices.

\begin{theorem}
In the $M$-ary hypothesis testing problem with deterministic likelihood ratio test decision rules, the minimax Bayes risk error divergence optimal decision weight $\mathbf{a}_k$ satisfies:
\begin{equation}
\label{eq:Mak}
	\nabla J(\mathbf{a}_k) = \sum_{i=1}^{v_k} w_i \nabla J(\mathbf{b}_{k,i}), 
\end{equation}
where the weights satisfy $w_i \ge 0$ and $\sum_i^{v_k} w_i = 1$.

Putting all of the $w_i$ into a vector $\mathbf{w}$, the optimal weight vector is the solution to the following optimization problem:
\begin{equation}
\label{eq:lambdaopt}
	\max_{\mathbf{w}} \sum_{i=1}^{v_k} w_i d\left(\nabla^{-1}J\left(\sum_{j=1}^{v_k} w_j \nabla J(\mathbf{b}_{k,j})\right)\Bigg\|\mathbf{b}_{k,i}\right)
\end{equation}
subject to the same constraints $w_i \ge 0$ and $\sum_i^{v_k} w_i = 1$.
\end{theorem}
\begin{IEEEproof}
The result follows by specializing \cite[Section 3]{NockN2005}, which applies to all Bregman divergences, to Bayes risk error divergence.
\end{IEEEproof}
The optimization problem \eqref{eq:lambdaopt} is similar to that solved in learning support vector machines \cite{NockN2005}.  The $w_i > 0$ that are found are `support' vertices that contribute to the location of the decision weight \cite{NockN2005}.

We note the centroid condition in the binary case $J'(a_k) = (J(b_k) - J(b_{k-1}))/(b_k - b_{k-1})$ \eqref{eq:ak} can be expressed as $J'(a_k) = w_1J'(b_{k-1}) + w_2J'(b_k)$, with $w_1,w_2 \ge 0$ and $w_1 + w_2 = 1$ due to the concavity of the Bayes risk and the intermediate value theorem of calculus.  In this form, we see the correspondence to \eqref{eq:Mak}.  In contrast to the $M$-ary case, there is a closed form expression for the decision weight $a_k$ in the binary case without requiring solving an optimization program.

\section{Rate--Distortion Analysis}
\label{sec:analysis}

To understand how quickly or slowly group minimax hypothesis testing approaches the performance of Bayesian hypothesis testing, in this section we examine the maximum achieved distortion as a function of the number of groups $K$.  Let us denote the minimax distortion overall as:
\begin{equation}
\label{eq:D}
		D = \min_{q_K}\max_{ {\bf p}} d({\bf p}\|q_K({\bf p})).
\end{equation}

\begin{theorem}
In the $M$-ary hypothesis testing problem with deterministic likelihood ratio test decision rules, the maximum Bayes risk error $D$ of the minimax-optimal quantization with $K$ groups satisfies the rate--distortion expression:
\begin{equation}
\label{eq:ratedistortion}
	K = O\left(\frac{1}{(M-1)!D^{\frac{M-1}{2}}}\right)
\end{equation}
\end{theorem}
\begin{IEEEproof}
The result follows from the fact that the volume of the probability simplex in the $M$-ary detection problem is $\frac{1}{(M-1)!}$ and specializing the results on $\epsilon$-nets for general Bregman divergences given in \cite[Lemma 14]{NielsenBN2007} to Bayes risk error divergence.
\end{IEEEproof}

The convergence from the edge case of minimax hypothesis testing to the other edge case of Bayesian hypothesis testing is in proportion to $K^{-2}$ in the binary hypothesis testing case, which is the same scaling seen in the mean Bayes risk error case presented in \cite{VarshneyV2008b}.  A similar scaling is also noted for detectors based on estimated prior probabilities \cite{JiaoZN2012}.  The minimax error scaling can be viewed as the asymptotic behavior of the minimum covering radius with respect to Bayes risk error divergence.  Note that all Bregman divergences, including squared error, will yield the same scaling behavior for $K$ as a function of $D$.  This implies that grouping by an incorrect Bregman fidelity criterion will incur a constant asymptotic rate loss.

\section{Examples}
\label{sec:examples}

We present two signal detection problem examples approached through group minimax hypothesis testing with optimal grouping.  The first example is the typical example of detecting a signal through Gaussian noise.  The second example is a ternary hypothesis testing problem with three different exponential likelihoods.

\subsection{Detecting Signals in Gaussian Noise}
\label{sec:gaussian}

Let us consider the following signal and measurement model:
\begin{equation}
\label{eq:scalarmeas}
	Y = s_m + W, \quad m \in \{0,1\},
\end{equation}
where $s_0=0$ and $s_1=\mu$, and $W$ is a zero-mean, Gaussian random variable with variance $\sigma^2$.  The parameters $\mu$ and $\sigma^2$ are known, deterministic quantities.  The error probabilities for this signal model are:
\begin{align*}
\label{eq:errors_gauss}
	p_E^{\text{I}}(a) &= Q\left( \tfrac{\mu}{2\sigma} + \tfrac{\sigma}{\mu}\ln\left(\tfrac{c_{10}a}{c_{01}(1-a)}\right)\right), \text{ and}\\
	p_E^{\text{II}}(a) &= Q\left( \tfrac{\mu}{2\sigma} - \tfrac{\sigma}{\mu}\ln\left(\tfrac{c_{10}a}{c_{01}(1-a)}\right)\right),
\end{align*}
where
\begin{displaymath}
	Q(\alpha) = \tfrac{1}{\sqrt{2\pi}} \int_\alpha^\infty e^{-x^2/2} dx.
\end{displaymath}
These error probabilities can be put together to obtain the Bayes risk error expression for this detection task:
\begin{equation}
\label{eq:Bayesrisk_gauss}
	J(p_0,a) = c_{10}p_0Q\left( \tfrac{\mu}{2\sigma} + \tfrac{\sigma}{\mu}\ln\left(\tfrac{c_{10}a}{c_{01}(1-a)}\right)\right) + c_{01}(1-p_0)Q\left( \tfrac{\mu}{2\sigma} - \tfrac{\sigma}{\mu}\ln\left(\tfrac{c_{10}a}{c_{01}(1-a)}\right)\right).
\end{equation}

We use the Lloyd--Max algorithm to design quantizers for the proposed group minimax criterion using the centroid and nearest neighbor conditions derived in Section \ref{sec:optimality}: equations \eqref{eq:ak} and \eqref{eq:bk}.  We show such quantizers for $K = 4$ and different ratios of the Bayes costs $c_{10}$ and $c_{01}$ along with different ratios of $\mu$ and $\sigma^2$.  As a point of comparison, we also show the optimal quantizers designed to minimize mean Bayes risk error divergence \cite{VarshneyV2008b}, rather than minimize maximum Bayes risk error divergence.  

\begin{figure}
	\begin{center}
		\begin{tabular}{c}
			\includegraphics[width=0.49\textwidth]{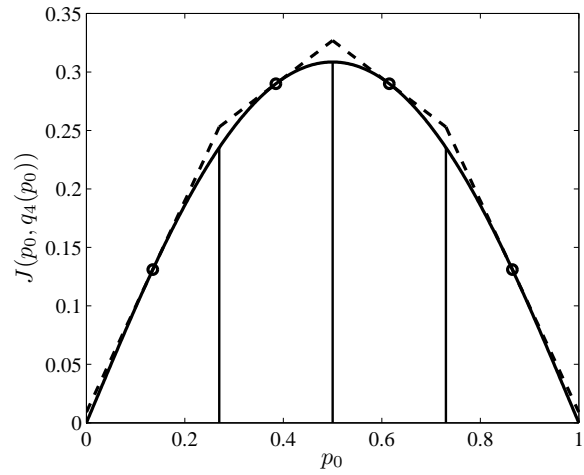} \\
			\footnotesize{(a)} \\
			\includegraphics[width=0.49\textwidth]{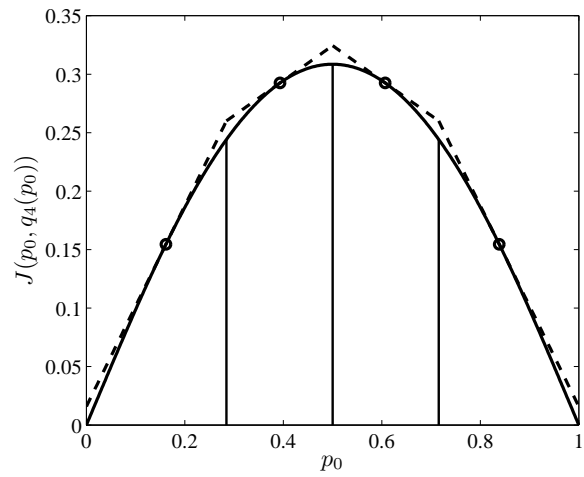} \\
			\footnotesize{(b)}
		\end{tabular}
	\end{center}
	\caption{(a) Minimum mean and (b) minimax Bayes risk error quantizers for $\mu = 1$, $\sigma^2 = 1$, $c_{10} = 1$, $c_{01} = 1$.}
	\label{fig:quant_1111}
\end{figure}

\begin{figure}
	\begin{center}
		\includegraphics[width=0.49\textwidth]{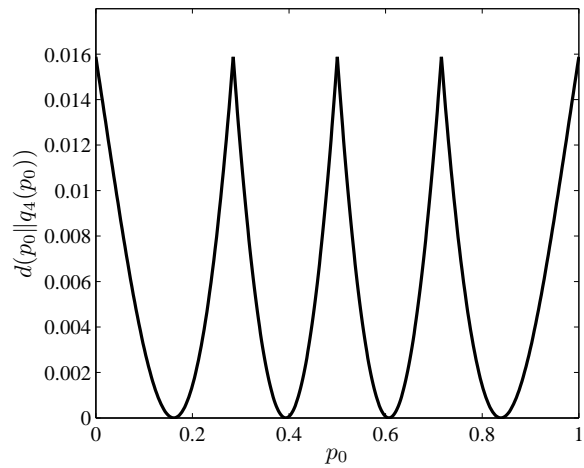}
	\end{center}
	\caption{Divergence of minimax quantizer for $\mu = 1$, $\sigma^2 = 1$, $c_{10} = 1$, $c_{01} = 1$.}
	\label{fig:d_1111}
\end{figure}

\begin{figure}
	\begin{center}
		\begin{tabular}{c}
			\includegraphics[width=0.49\textwidth]{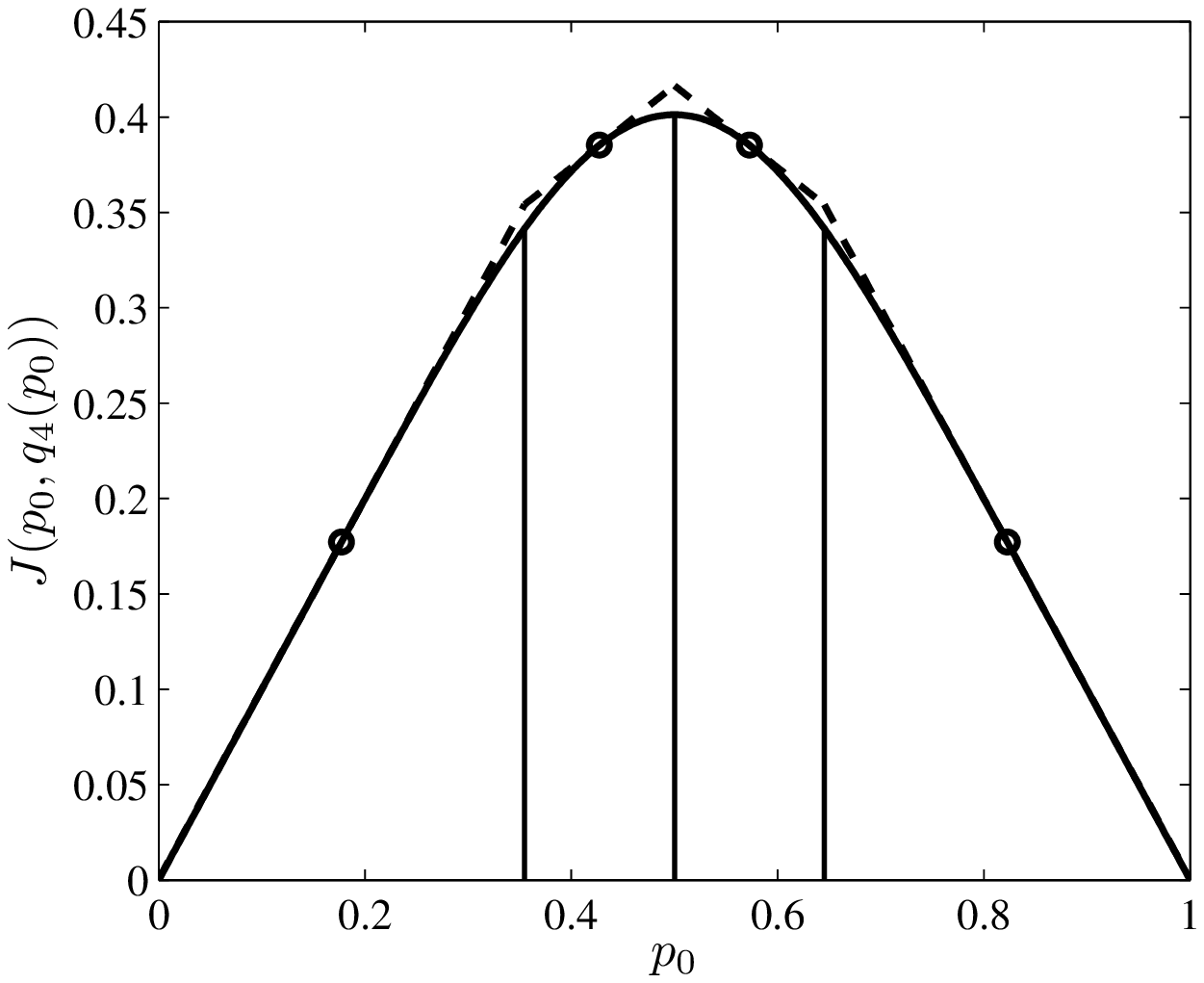} \\
			\footnotesize{(a)} \\
			\includegraphics[width=0.49\textwidth]{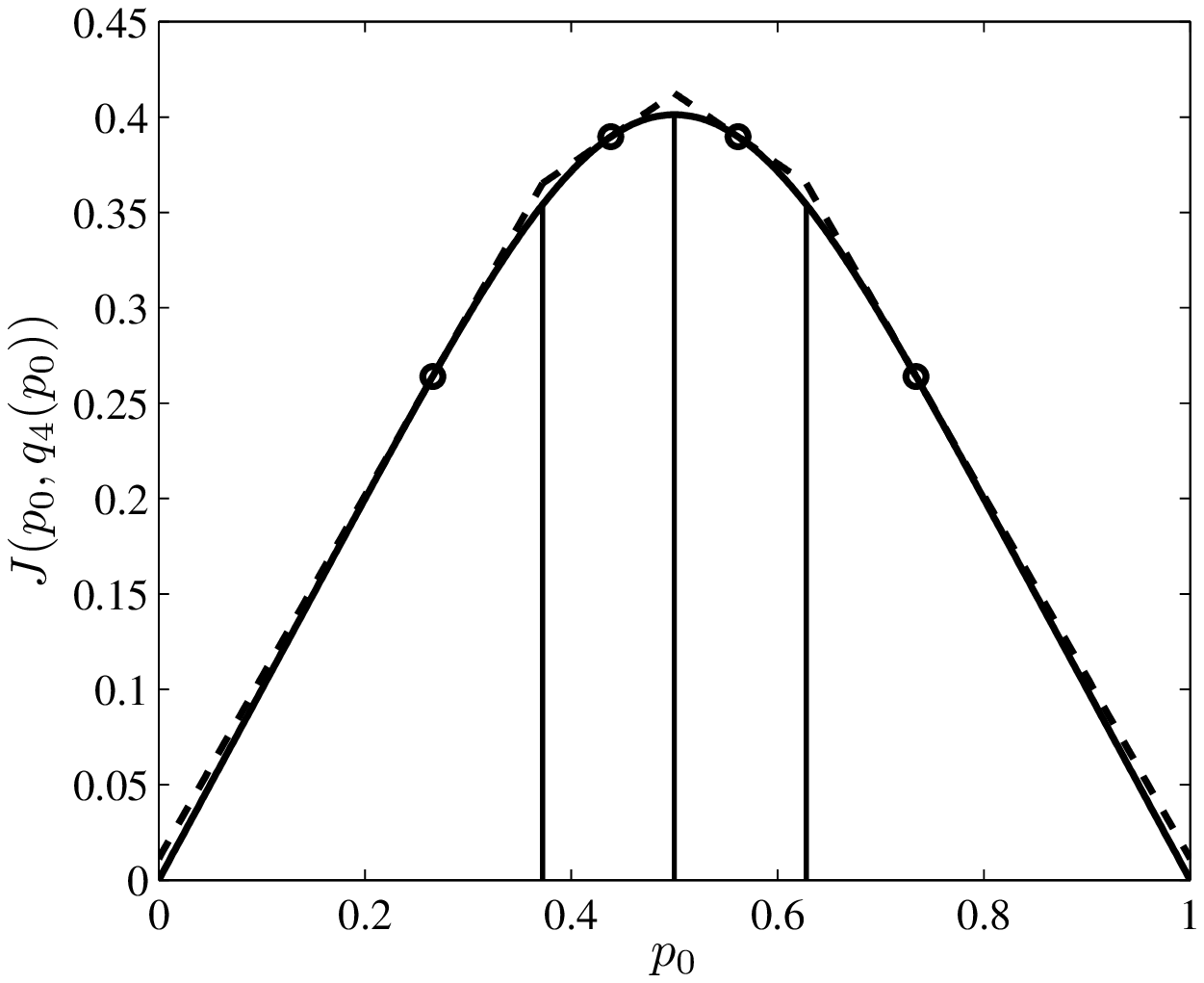} \\
			\footnotesize{(b)}
		\end{tabular}
	\end{center}
	\caption{(a) Minimum mean and (b) minimax Bayes risk error quantizers for $\mu = 1$, $\sigma^2 = 2$, $c_{10} = 1$, $c_{01} = 1$.}
	\label{fig:quant_1211}
\end{figure}

\begin{figure}
	\begin{center}
		\includegraphics[width=0.49\textwidth]{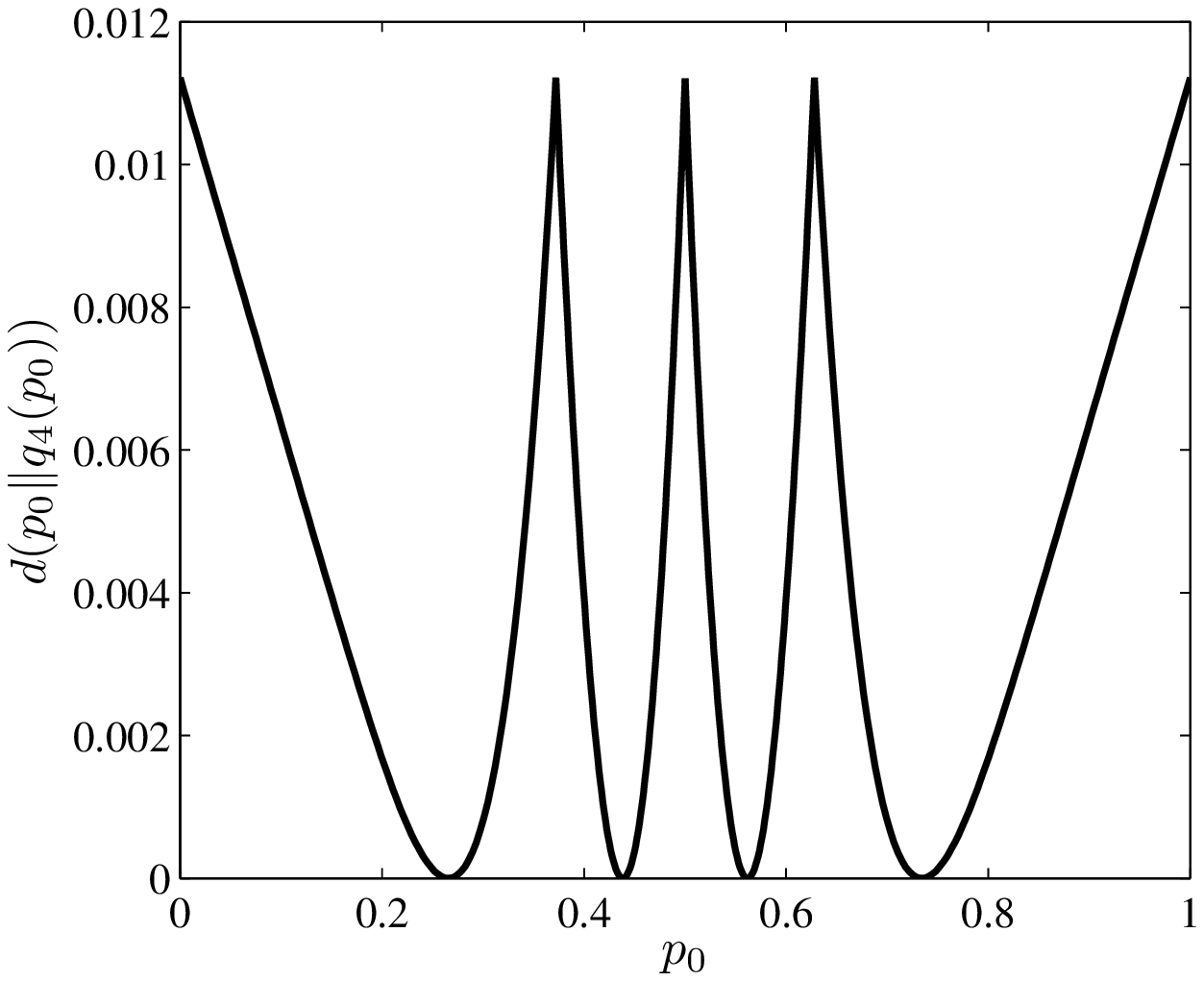}
	\end{center}
	\caption{Divergence of minimax quantizer for $\mu = 1$, $\sigma^2 = 2$, $c_{10} = 1$, $c_{01} = 1$.}
	\label{fig:d_1211}
\end{figure}

\begin{figure}
	\begin{center}
		\begin{tabular}{c}
			\includegraphics[width=0.49\textwidth]{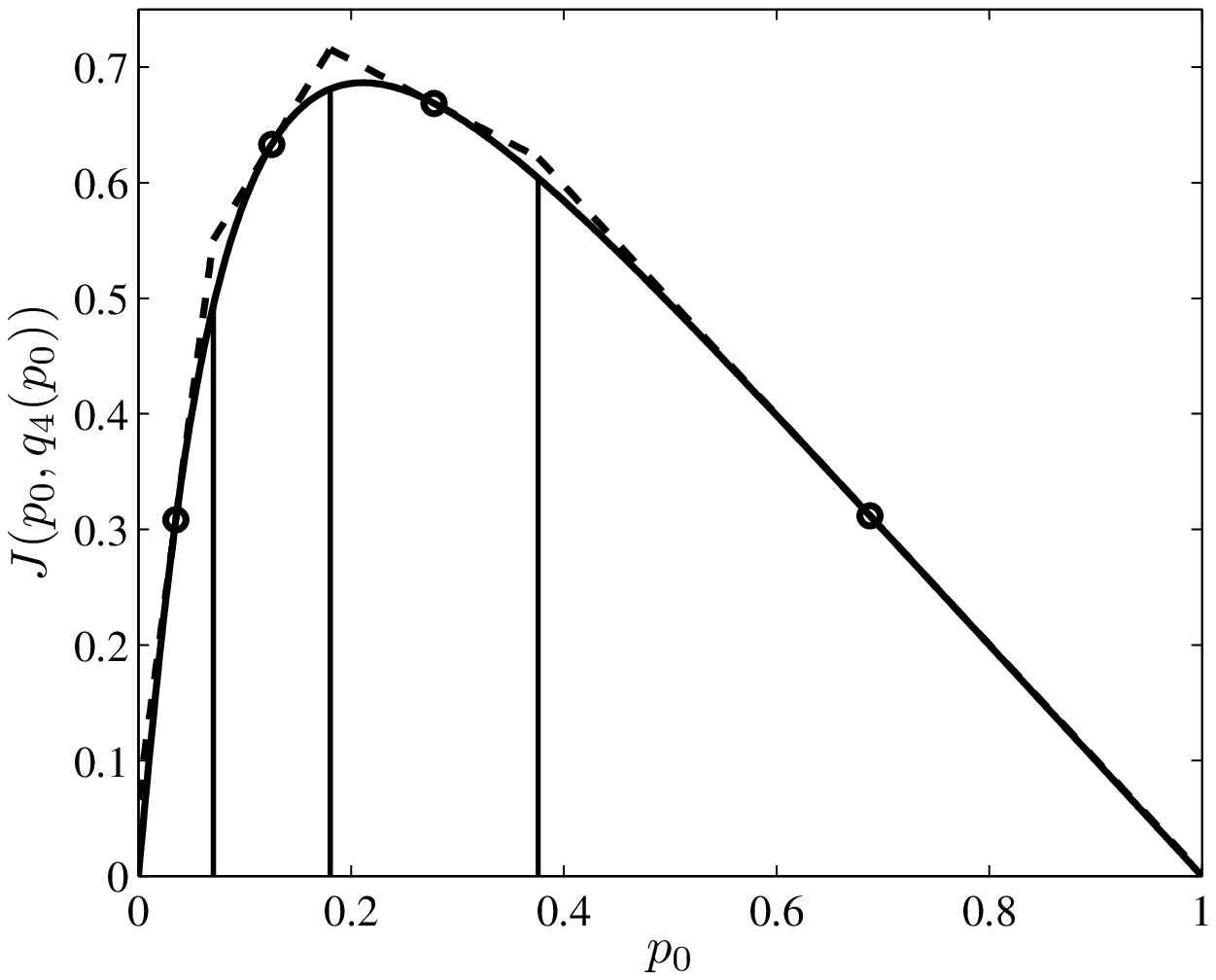} \\
			\footnotesize{(a)} \\
			\includegraphics[width=0.49\textwidth]{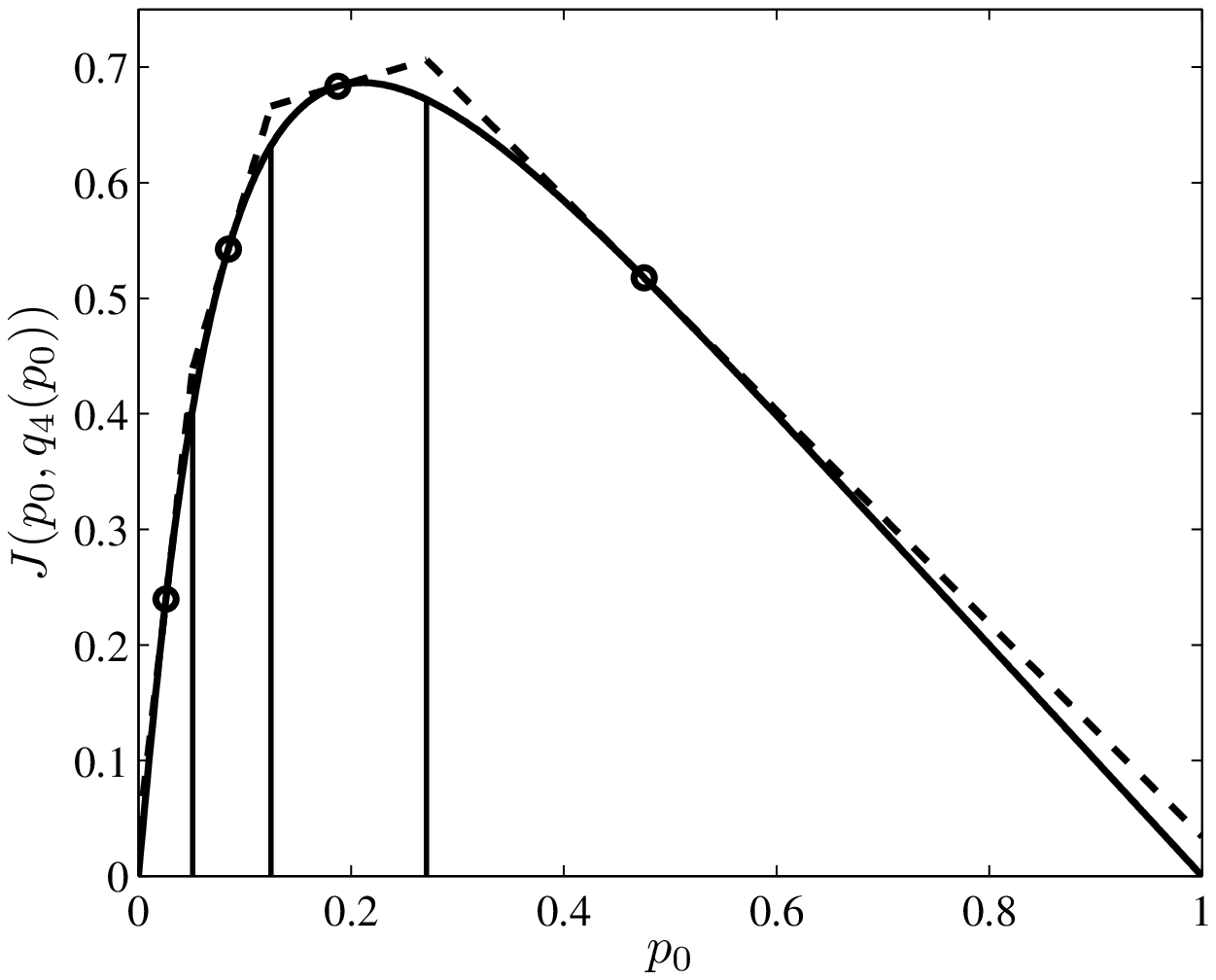} \\
			\footnotesize{(b)}
		\end{tabular}
	\end{center}
	\caption{(a) Minimum mean and (b) minimax Bayes risk error quantizers for $\mu = 1$, $\sigma^2 = 1$, $c_{10} = 10$, $c_{01} = 1$.}
	\label{fig:quant_11ten1}
\end{figure}

\begin{figure}
	\begin{center}
		\includegraphics[width=0.49\textwidth]{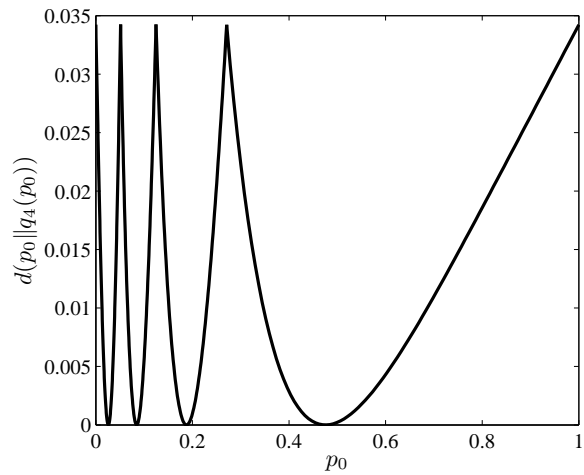}
	\end{center}
	\caption{Divergence of minimax quantizer for $\mu = 1$, $\sigma^2 = 1$, $c_{10} = 10$, $c_{01} = 1$.}
	\label{fig:d_11ten1}
\end{figure}

Fig.~\ref{fig:quant_1111} shows quantizers for equal Bayes costs and equal mean and standard deviation.
In the plots, the black curve is $J(p_0)$ and the dashed line is $J(p_0,q_4(p_0))$, with their difference being $d(p_0\|q_4(p_0))$.  The circle markers are the representation points and the vertical lines indicate the interval boundaries of the groups.  The divergence value $d(p_0\|q_4(p_0))$ is shown in Fig.~\ref{fig:d_1111}.  
The minimax groups and representation points are more clustered in the middle of the probability simplex and around the peak of $J(p_0)$ than the minimum mean groups and representation points.  This is more apparent in the quantizers for the noisier measurement model with $\mu = 1$ and $\sigma^2 = 2$ shown in Fig.~\ref{fig:quant_1211}, and the quantizers for unequal Bayes costs $c_{10} = 10$ and $c_{01} = 1$ shown in Fig.~\ref{fig:quant_11ten1}. The divergence values for these other two cases are shown in Fig.~\ref{fig:d_1211} and Fig.~\ref{fig:d_11ten1}.

\begin{figure}
	\begin{center}
		\begin{tabular}{c}
			\includegraphics[width=0.49\textwidth]{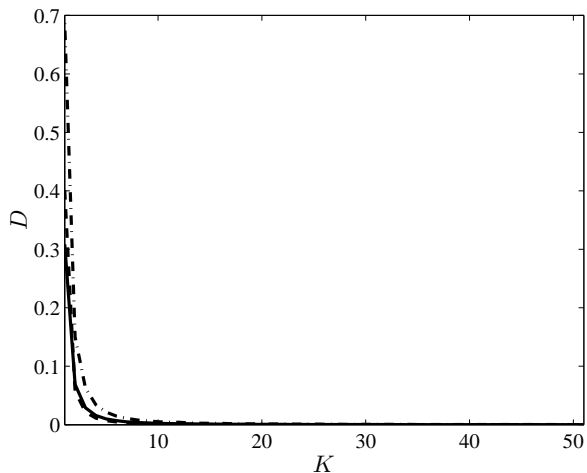} \\
			\footnotesize{(a)} \\
			\includegraphics[width=0.49\textwidth]{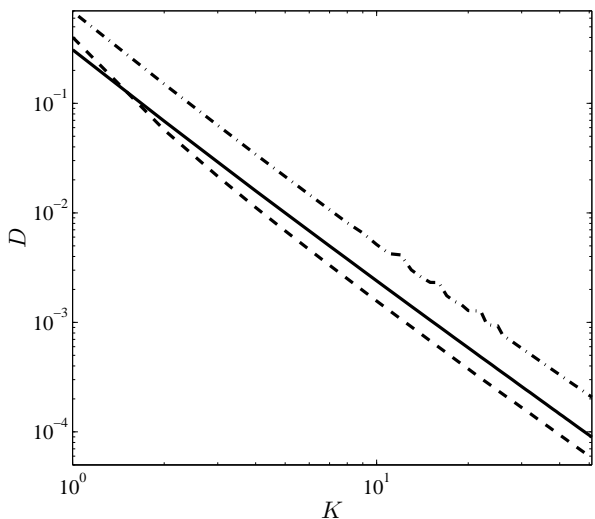} \\
			\footnotesize{(b)}
		\end{tabular}
	\end{center}
	\caption{Minimax Bayes risk error for $\mu = 1$, $\sigma^2 = 1$, $c_{10} = 1$, $c_{01} = 1$ (solid line), $\mu = 1$, $\sigma^2 = 2$, $c_{10} = 1$, $c_{01} = 1$ (dashed line), and $\mu = 1$, $\sigma^2 = 1$, $c_{10} = 10$, $c_{01} = 1$ (dashed and dotted line), on (a) linear and (b) logarithmic scales.}
	\label{fig:D}
\end{figure}

The minimax Bayes risk error as a function of $K$ for this example is shown in Fig.~\ref{fig:D} on both linear and logarithmic scales.
The curves seen in Fig.~\ref{fig:D}(b) exactly reflect the behavior expected according to the rate--distortion analysis of Section~\ref{sec:analysis}.  They are almost perfectly linear beyond a couple of small $K$ values.  The slopes of the lines are $-2$ which is the rate predicted for $M = 2$.

\subsection{Distinguishing Exponential Likelihoods}
\label{sec:exponential}

In this example, we consider objects in a queuing that are served at varying rates.  Objects are served at rate $\lambda_m > 0$ when in state $H = h_m$ for $m \in \{0,1,2\}$, with $\lambda_0 > \lambda_1 > \lambda_2$.  The ternary hypothesis testing task is to determine which state the object is in based on an observation of the time $Y = y$ at which it is served.  The likelihood functions take the form:
\begin{equation}
\label{eq:explik}
	f_{Y|H}(y|H = h_m) = \lambda_m e^{-\lambda_m y}.
\end{equation}
For simplicity, we only consider the case in which $c_{00} = c_{11} = c_{22} = 0$, and $c_{01} = c_{02} = c_{10} = c_{12} = c_{20} = c_{21} = 1$.

Recall that we denote the prior probabilities $p_0$ and $p_1$ through the vector $\mathbf{p}$ and the decision weights $a_0$ and $a_1$ through the vector $\mathbf{a}$ (where $p_2 = 1 - p_0 - p_1$ and $a_2 = 1 - a_0 - a_1$).  In this example, if we define the following two functions of the decision weights:
\begin{align}
\label{eq:gamma01}
	\gamma_{01}(\mathbf{a}) &= \max\left\{0,\frac{1}{\lambda_0 - \lambda_1}\ln\left(\frac{a_0\lambda_0}{a_1\lambda_1}\right)\right\}, \\
\label{eq:gamma12}
	\gamma_{12}(\mathbf{a}) &= \max\left\{0,\frac{1}{\lambda_1 - \lambda_2}\ln\left(\frac{a_1\lambda_1}{(1 - a_0 - a_1)\lambda_2}\right)\right\},
\end{align}
then the mismatched Bayes risk function is:
\begin{equation}
\label{eq:expoBayesrisk}
	J(\mathbf{p},\mathbf{a}) = p_0e^{-\lambda_0\gamma_{01}(\mathbf{a})} + p_1\left(1 - e^{-\lambda_1\gamma_{01}(\mathbf{a})} + e^{-\lambda_1\gamma_{12}(\mathbf{a})}\right) + (1 - p_0 - p_1)\left(1 - e^{-\lambda_2\gamma_{12}(\mathbf{a})}\right).
\end{equation}
We calculate the gradient $\nabla J(\mathbf{a})$ in closed form, but omit it here because of its unwieldy nature.

\begin{figure}
	\begin{center}
		\includegraphics[width=0.49\textwidth]{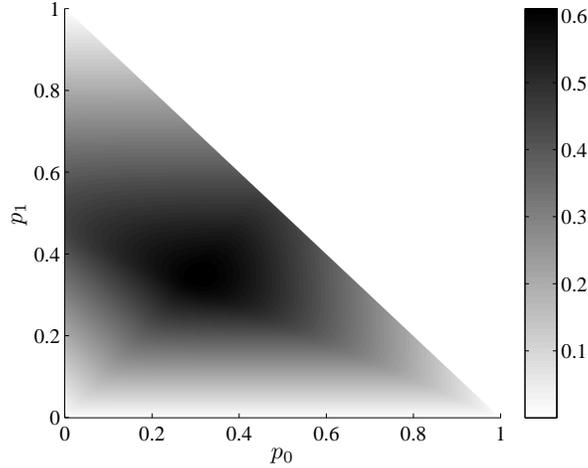}
	\end{center}
	\caption{Bayes risk function $J(\mathbf{p})$ for $\lambda_0 = 5$, $\lambda_1 = 4$, $\lambda_2 = 3$.}
	\label{fig:ternary_exponentials_bayesrisk}
\end{figure}

\begin{figure}
	\begin{center}
		\includegraphics[width=0.49\textwidth]{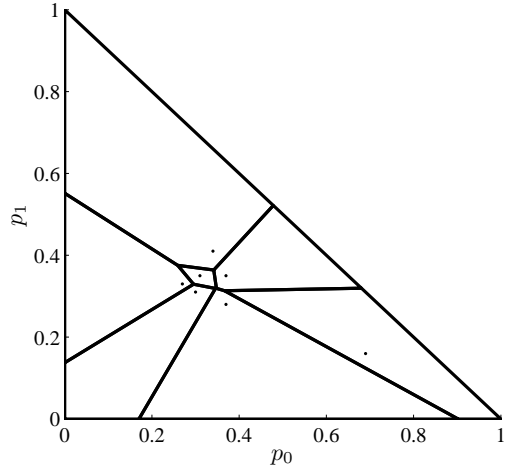}
	\end{center}
	\caption{Optimal groupings and representation points found for $\lambda_0 = 5$, $\lambda_1 = 4$, $\lambda_2 = 3$ with $K=7$.}
	\label{fig:ternary_exponentials_7}
\end{figure}

\begin{figure}
	\begin{center}
		\includegraphics[width=0.49\textwidth]{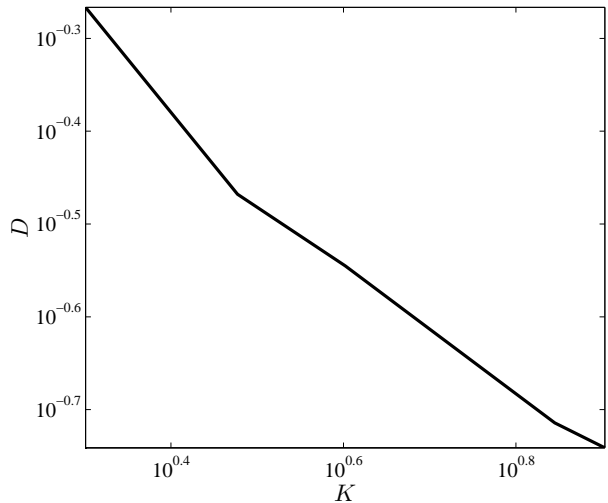}
	\end{center}
	\caption{Minimax Bayes risk error for $\lambda_0 = 5$, $\lambda_1 = 4$, $\lambda_2 = 3$  on a logarithmic scale.}
	\label{fig:D3}
\end{figure}

We now examine optimal groupings for group minimax hypothesis testing in the ternary exponential service time example with $\lambda_0 = 5$, $\lambda_1 = 4$, and $\lambda_2 = 3$.  The convex Bayes risk function defined over the probability simplex is shown via shading in Fig.~\ref{fig:ternary_exponentials_bayesrisk}.
The Bayes risk function is zero at all three corners and along the $p_0$ axis.  We apply the alternating nearest neighbor condition and centroid condition of the Lloyd--Max algorithm to find the optimal groupings in the $K=7$ case.  Fig.~\ref{fig:ternary_exponentials_7} is a plot of the groups and representation points that are found.
In Fig.~\ref{fig:D3}, we show the minimax error for this example as a function of $K$ in the logarithm-transformed domain.  As expected for $M = 3$, the function is approximately linear with slope $-1$.

\section{Conclusion}
\label{sec:conclusion}

The group minimax test---as an intermediate formulation between the Bayesian and minimax tests that takes advantage of set-structured, incomplete advanced knowledge of priors---was proposed long ago by the early decision theorists.  However results in the literature were obtained under special circumstances and when the sets were predetermined.  In this work, we approach group minimax through the emergent theory of quantizing with Bregman divergences and make statements about optimal representative priors that do not rely on any special likelihood functions.  By optimizing the minimax Bayes risk error divergence, we obtain a closed-form Stolarsky mean expression for the optimal representative prior within a group in the binary case.  In the $M$-ary case, we present a support vector machine-like program to be solved.

In descriptions of group minimax or $\Gamma$-minimax in the literature, no heed is given to determining the best $K$ groups to maximize detection performance.  We solve this problem jointly with finding representative priors within groups through an alternating minimization involving Bregman centroids and Bregman bisectors.  The optimal groupings are delineated by a Voronoi diagram or $\epsilon$-net of the space of prior probabilities.  We give closed-form expressions for the polyhedral group boundaries.  Moreover, in a rate--distortion format, we characterize the rate at which detection performance of group minimax approaches Bayesian detection as the number of optimal groups increases.

The research described in this paper is for single decision makers.  Distributed detection with multiple agents working as a team \cite{RhimVG2012,RhimVG2013,GulZ2013} or with conflicts \cite{GruenwaldD2004,RhimVG2011b} can also be considered.  Additionally, regret theory is closely connected with minimax hypothesis testing \cite{LoomesS1982,EldarBN2004}; extensions of this paper within the confines of regret theory may be explored.

\section*{Acknowledgment}

The authors thank Joong Bum Rhim for discussions.

\ifCLASSOPTIONcaptionsoff
  \newpage
\fi

\bibliographystyle{IEEEtran}
\bibliography{IEEEabrv,jminimax}

\end{document}